\documentclass{aa}

\usepackage{graphicx}
\usepackage[varg]{txfonts}
\usepackage{amsmath}
\usepackage{amssymb}
\usepackage{textcomp}
\usepackage{natbib}
\bibpunct{(}{)}{;}{a}{}{,}

\begin{document}
%\linenumbers
\title{A Highly Magnetized Twin-Jet Base Pinpoints a Supermassive Black Hole} 
\author{%
  A.-K. Baczko\inst{\ref{affil:remeis},\ref{affil:wuerzburg},\ref{affil:mpifr},*}
  \and R. Schulz\inst{\ref{affil:wuerzburg},\ref{affil:remeis},\ref{affil:astron}}
  \and M. Kadler\inst{\ref{affil:wuerzburg}}
  \and E. Ros\inst{\ref{affil:mpifr},\ref{affil:dep_valencia2},\ref{affil:dep_valencia}}
  \and M. Perucho\inst{\ref{affil:dep_valencia},\ref{affil:dep_valencia2}}
  \and T.\,P. Krichbaum\inst{\ref{affil:mpifr}}
  \and M. B\"ock\inst{\ref{affil:mpifr}}
  \and M. Bremer\inst{\ref{affil:idrm}}
  \and C. Grossberger\inst{\ref{affil:mpife},\ref{affil:wuerzburg},\ref{affil:remeis}}
  \and M. Lindqvist\inst{\ref{affil:onsala}}
  \and A.\,P. Lobanov\inst{\ref{affil:mpifr},\ref{affil:hamburg}}
  \and K. Mannheim\inst{\ref{affil:wuerzburg}}
  \and I. Mart{\'{\i}}-Vidal\inst{\ref{affil:onsala}}
  \and C. M\"uller\inst{\ref{affil:nijmegen},\ref{affil:wuerzburg},\ref{affil:remeis}}
  \and J. Wilms\inst{\ref{affil:remeis}}
  \and J.\,A. Zensus\inst{\ref{affil:mpifr}}
    }%
\institute{%
 Dr. Remeis-Sternwarte \& ECAP, Universit\"at Erlangen-N\"urnberg, Sternwartstrasse 7, 96049 Bamberg, Germany \label{affil:remeis}
 \and
 Institut f\"ur Theoretische Physik und Astrophysik, Universit\"at W\"urzburg, Emil-Fischer-Strasse 31, 97074 W\"urzburg, Germany\label{affil:wuerzburg}
 \and
 Max-Planck-Institut f\"ur Radioastronomie, Auf dem H\"ugel 69, 53121 Bonn, Germany \label{affil:mpifr}
 \and
 Observatori Astron\`omic, Universitat de Val\`encia, C/ Catedr\'atico Jos\'e Beltr\'an no. 2, 46980 Paterna, Val\`encia, Spain \label{affil:dep_valencia2}
 \and
 Departament d'Astronomia i Astrof\'isica, Universitat de Val\`encia, C/ Dr. Moliner 50, 46100 Burjassot, Val\`encia, Spain \label{affil:dep_valencia}
 \and
 Institut de Radioastronomie Millim\'etrique, 300 rue de la Piscine, Domain Universitaire, 38406 Saint Martin d'H\`eres, France \label{affil:idrm}
 \and
 Max-Planck-Institut f\"ur extraterrestrische Physik, Giessenbachstrasse 1, 85748 Garching, Germany\label{affil:mpife}
 \and
 Department of Earth and Space Sciences, Chalmers University of Technology, Onsala Space Observatory, 439 92 Onsala, Sweden\label{affil:onsala}
 \and
 Institut f\"ur Experimentalphysik, Universit\"at Hamburg, Luruper Chaussee 149, 22761 Hamburg, Germany\label{affil:hamburg}
 \and
 Department of Astrophysics/IMAPP, Radboud University Nijmegen, PO Box 9010, 6500 GL, Nijmegen, The Netherlands\label{affil:nijmegen}
 \and
 Netherlands Institute for Radio Astronomy (ASTRON), PO Box 2, 7990 AA Dwingeloo, The Netherlands\label{affil:astron}
 }%
  
\date{}

\abstract{
Supermassive black holes (SMBH) are essential for the production of jets in radio-loud active galactic nuclei (AGN). Theoretical models based on Blandford \& Znajek extract the rotational energy from a Kerr black hole, which could be the case for \object{\object{NGC\,1052}}, to launch these jets. This requires magnetic fields  of the order of $10^3\,$G to $10^4\,$G. We imaged the vicinity of the SMBH of the AGN \object{\object{NGC\,1052}} with the Global Millimetre VLBI Array and found a bright and compact central feature, smaller than 1.9 light days (100 Schwarzschild radii) in radius. Interpreting this as a blend of the unresolved jet bases, we derive the magnetic field at 1 Schwarzschild radius to lie between 200 G and $\sim 8\times10^4$ G consistent with Blandford \& Znajek models.
}
\titlerunning{A Highly magnetized Twin-Jet Base Pinpoints a SMBH}
\maketitle

\section{Introduction}

Together with the existence of a central supermassive black hole surrounded by a circumnuclear torus, the unification scheme of radio loud AGN \citep{Ant93, Fan11} postulates the occurrence of two symmetric radio jets, which emanate parallel and anti-parallel to the angular momentum vector of the black hole. 
Despite our expectations, we mostly find jets appear one-sided in radio flux-density-limited samples of AGN such as MOJAVE  \citep{Lis09}. This is a selection effect due to the strongly Doppler-boosted emission of the relativistically moving plasma in those jets, which are pointed at small angles to the line of sight towards the observer \citep{Lis97}.
This orientation bias poses a problem when trying to locate the putative black hole at the true center of activity because of the unknown offset between the black hole and the  base of the jet above the accretion disk. 
Following \cite{Bla77} it is possible to launch the jets by extracting rotational energy from a Kerr black hole. To test whether high magnetic fields as needed for this approach are observable, a precise knowledge of the location of the central engine is required. 
Ultra-high-resolution radio-interferometric observations with Very Long Baseline Interferometry (VLBI) have the potential to locate regions in jets with microarcsecond precision \citep{Zen97}.

The core shift effect has been studied in detail in the low-luminosity AGN \object{NGC\,1052} (RA 02h41m04.8s, DEC$-$08d15'21'')\footnote{NASA/IPAC Extragalactic Database (NED) and references there} at cm wavelengths by  \citet{Kad04b}. The source hosts a supermassive black hole with a mass of $\sim$\,$10^{8.2}M_\sun$  \citep{Woo02} and a twin-jet system. It is located at a distance of 19.5\,Mpc\addtocounter{footnote}{-1}\footnotemark. 
A two-sided radio structure stretches across the inner $\sim 3$\,kpc of the galaxy, interacting strongly with the interstellar medium \citep{Wro84,Kad04b}, from which a total jet kinetic power of $5 \times 10^{41}$erg\,s$^{-1}$ is estimated \citep{Kad03}.
The parsec-scale structure of the twin-jet system has been studied extensively at cm-wavelengths using multi-frequency VLBI observations \citep{Kam01,Ver03,Kad04a}, revealing a prominent emission gap between the two jets, which has been interpreted as being due to free-free absorption in a circumnuclear torus with a column density of $N_\mathrm{H}\sim 10^{22}$--$10^{24}$\,cm$^{-2}$  \citep{Kam01,Ver03,Kad04a,Kad04b,Bre09} around a probably rapidly rotating black hole \citep{Bre09}.

The free-free absorption effects that dominate the morphology of the source at cm wavelengths \citep{Kam01,Ver03,Kad04a}, are negligible at 3\,mm (corresponding to a frequency of 86\,GHz). Therefore, observations at 86\,GHz are able to peer through the absorbing structure and reveal insights into the jet-launching region of NGC\,1052, an exceptional case of a double-sided jet system.

In this work, we present results from one observation of the twin-jet system of \object{NGC\,1052} at 86\,GHz with the Global mm-VLBI Array (GMVA\footnote{\url{http://www3.mpifr-bonn.mpg.de/div/vlbi/globalmm/}}) in 2004, yielding an unobscured view of the innermost structures in the twin-jet system, where 1\,mas corresponds to a projected linear size of 0.094\,pc.

VLBI observations at such a high frequency are a challenge for weak and low-declination sources such as \object{NGC\,1052}, especially in terms of sensitivity. Present GMVA capabilities offer a large number of telescopes capable to observe at 86\,GHz and much enhanced data bit rates, leading to high enough visibility sampling to image the twin-jet system with the highest resolution until now.
We discuss the data reduction in the next section and report the results and conclusions from our successful calibration and imaging in Sections 3 to 5.

\section{Observation and Data Reduction}
\label{Observation_Reduction}

\begin{figure}[!h]
\includegraphics[width=1\linewidth]{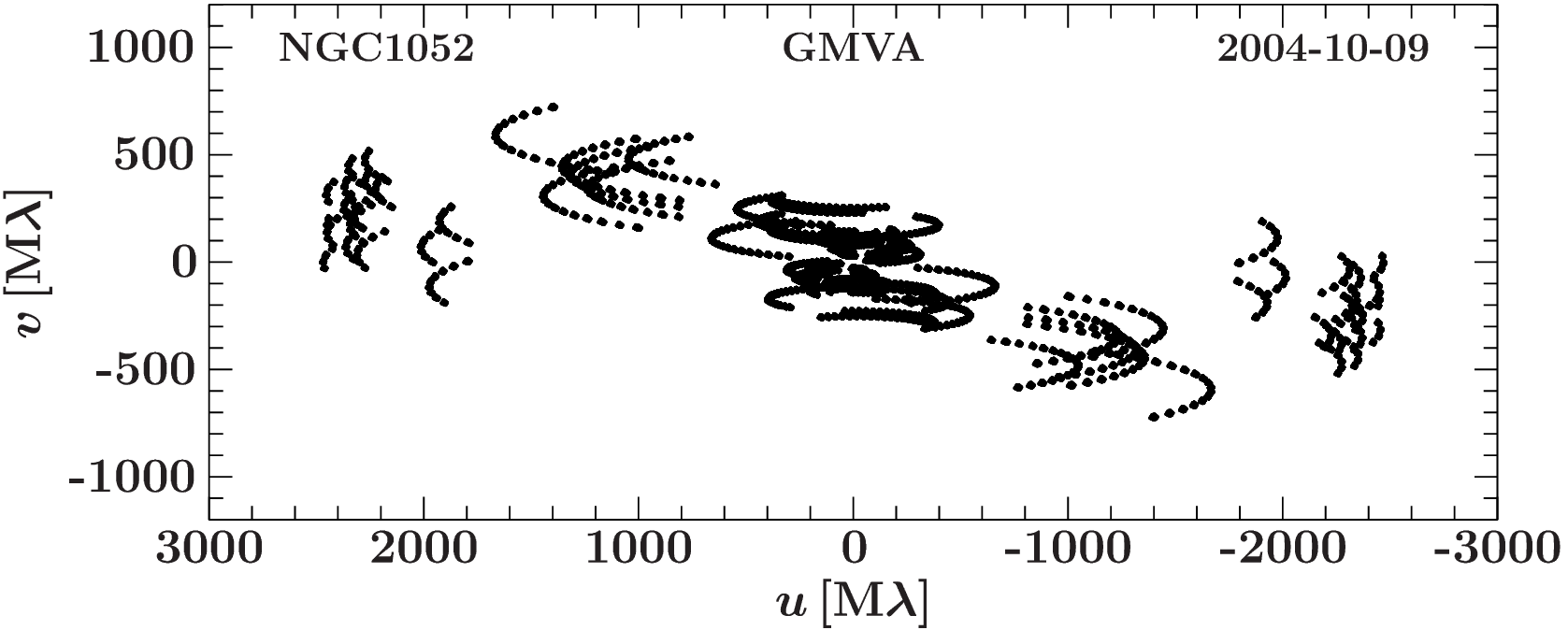}
\caption{$(u,v)$-coverage of \object{NGC\,1052} at 86\,GHz with the GMVA.}
\label{fig:uvpl}
\end{figure}

The observations were performed during the Session II (October) of the GMVA in 2004 in experiment GK027, from October 09 to 10 at a frequency of 86 GHz with 15 antennas (see Table~\ref{tab:GMVA}). The European antennas started on October 9 at 22:00 UT and ended at 07:00 UT, whereas the US array observed from 02:00 to 14:00 UT. 
The correlator accumulation time was 1s, the data were recorded in setup mode 512-8-2, namely, with eight 16-MHz channels in left-hand circular polarization. The observing cycle was several blocks of 20 min, consisting in one 1-min scan on the calibrator \object{0235+164}, 2 min for slewing, by 3m24s on NGC 1052, and 13.5 min for pointing and calibration and slewing back to \object{0235+164}. \object{0716+714} was observed as calibrator immediately prior to the \object{NGC\,1052} block, and \object{OJ\,287} was observed at the end.
Data were recorded with the Mark\,IV and Mark\,V systems and correlated at the processing center at the Max-Planck-Institut f\"ur Radioastronomie, Germany. The final $(u,v)$-coverage is shown in Fig.\ref{fig:uvpl} and reaches maximum values of $\sim 2.2$\,G$\lambda$ in the east-west direction.

\begin{table}[!h]
	\centering
	\caption[]{List of participating GMVA stations.}
	\label{tab:GMVA}
	\begin{tabular}{cccc}
	  \hline\hline
	  Station & Diameter 	& SEFD& Observation\\
		  &    [m]	& [Jy]&	\\\hline
	  Mets\"{a}hovi     & 14 & 17650 & \checkmark \\
	  Onsala 	    & 20 & 5100 & \textsf{--}\tablefootmark{a}\\
	  Effelsberg 	    & 100& 930  & \checkmark\\
	  Plateau de Bure   & 35\tablefootmark{b} & 450  & \checkmark\\
	  Pico Veleta 	    & 30 & 640  & \checkmark\\
	  VLBA NL	    & 25 & 4550 & \checkmark\\
	  VLBA FD	    & 25 & 3030 & \checkmark\\
	  VLBA LA	    & 25 & 1390 & \checkmark\\
	  VLBA KP	    & 25 & 3450 & \checkmark\\
	  VLBA PT	    & 25 & 2270 & \checkmark\\
	  VLBA OV	    & 25 & 3570 & \checkmark\\
	  VLBA BR	    & 25 & 3230 & \checkmark\\
	  VLBA MK	    & 25 & 5560 & \checkmark\\
	\hline		
	\end{tabular}
	\tablefoot{
	\tablefoottext{a}{data loss due to receiver problems.}
	\tablefoottext{b}{Equivalent ($5\times15$m).}
}
\end{table}

Amplitude and phase calibration were performed using the NRAO Astronomical Image Processing System (AIPS). After correcting for changes in the parallactic angle the phase calibration was performed in two steps. First, the {sub-band phase calibration} corrects for station dependent delays and phase offsets with each sub-band due to differences in the electronic systems of each station \citep{Mar12a}. In contrast to a manual phase and delay calibration we used an {interpolated} smoothed solution to align the phases of all subbands, also referred to as intermediate frequencies (IFs). This solution was found by interpolating the subset of best phase and delay solutions for each individual IF referred to one fiducial IF. The {multi-band fringe fitting} corrects for the multi-band antenna-related gains with the use of global fringe fitting.
For the fringe search, we used the AIPS task FRING. We selected data above a SNR threshold of 4.5 (aparm(7) in FRING). The search windows were of 20 ns, and 100 mHz, for delay and delay rate, respectively (parameters dparm(2) and dparm(3), respectively).  After several tests, we found that the highest number of successful detections was obtained with an integration time of 1 min (solint parameter in FRING). This is a good trade-off between phase degradation due to coherence times of 20-30 seconds and detection as a function of the square root of the integration time. We used Pico Veleta and Los Alamos as reference antennas, and we applied exhaustive fringe search (aparm(9)=1) first to the Effelsberg and then Pie Town stations. Delay and rate solutions with a very clear departure from the average solutions for each antenna were flagged after visual inspection.
The amplitude calibration was made using weather data recorded at each station and measurements of the system temperatures at all observing telescopes. The used task was \textsc{apcal}. After calibration, the phases are flat and the amplitude can be expressed in units of jansky (prior to calibration those are simply correlation coefficients).

\begin{figure}[!h]
\includegraphics[width=1\linewidth]{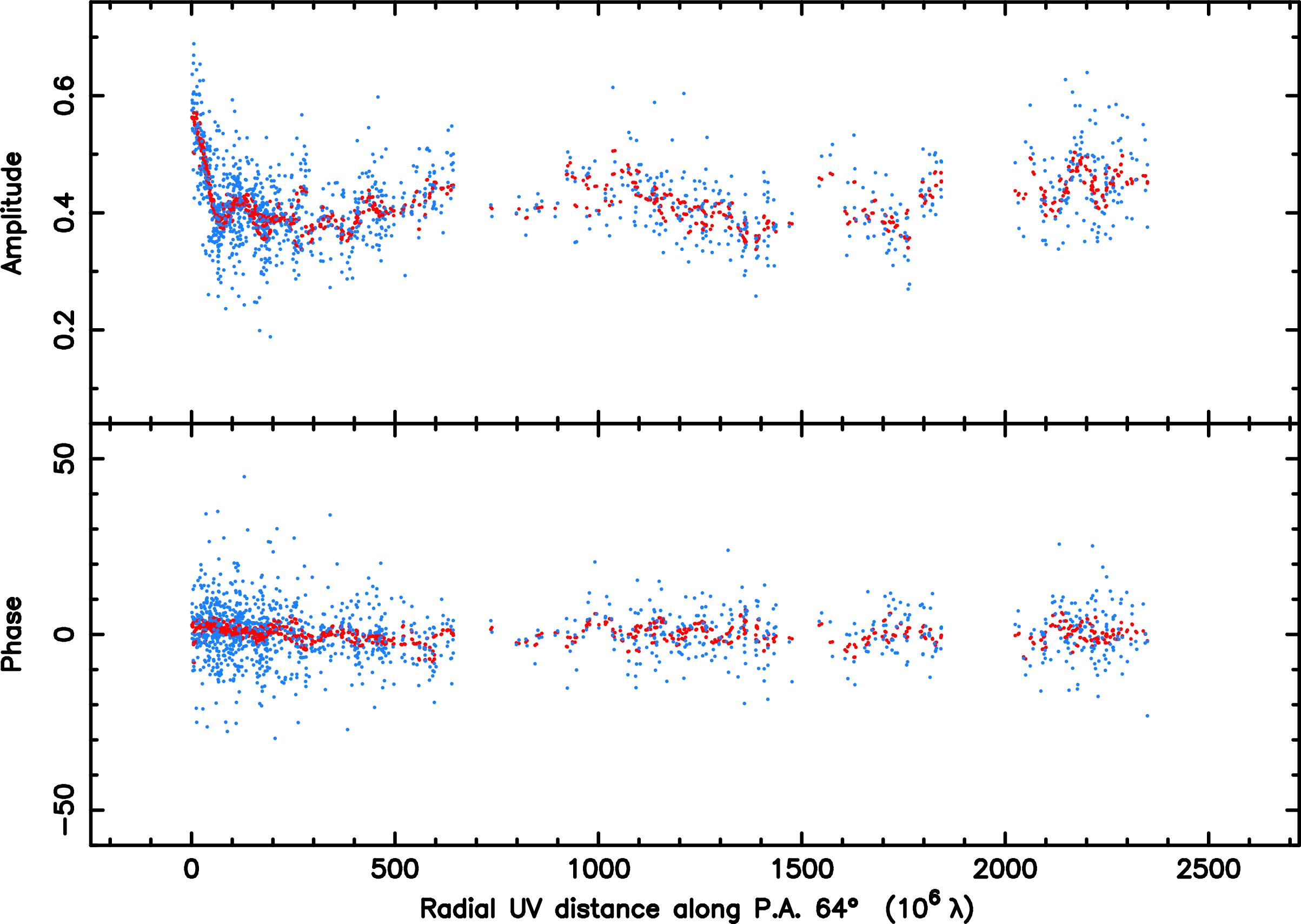}
\caption{
Visibility amplitude (top panel) and phase (lower panel) versus $(u,v)$-distance of \object{NGC\,1052} at 86\,GHz with the GMVA. The visibility data shown are uniformly weighted, post-self-calibrated ones, averaged per scan (240\,s) and projected along the jet positional angel of $64^\circ$. The blue and red dots correspond to the visibility data and the clean model, respectively.}
\label{fig:radplot}
\end{figure}
After the calibration steps in AIPS, hybrid mapping with the DIFMAP software \citep{She94} was performed. The data were re-binned to 10\,s, and statistical errors were determined from the scatter within these bin periods for each baseline. The visibilities were carefully inspected and bad data points were flagged. After building a first basic model using the \textsc{CLEAN} algorithm implemented in DIFMAP and phase self-calibration, the data were edited again. Once a good model was found, the visibilities were amplitude self-calibrated over the complete observation time. The process was repeated for subsequently smaller solution intervals of amplitude self-calibration. 

We applied uniform weighting of the visibilities to achieve the best angular resolution. The imaging process was repeated several times to check for ambiguities in the final \textsc{CLEAN}-model. Based on this and the amplitude of calibration factors, we estimate a conservative error of 15 \% on the flux densities in the final image. The image parameters are listed in Table~\ref{tab:ImageParameters}. 
Figure~\ref{fig:radplot} shows the visibility amplitude and phase versus radial $(u,v)$-distance in blue with the final clean model overplotted in red.

\begin{table*}
	\centering
	\caption[]{Image Parameters of the GMVA observation from 2004-10-09.}
	\label{tab:ImageParameters}
	\begin{tabular}{@{}cccccccc@{}}
	\hline\hline 
	Weighting & Taper & $S_\mathrm{total}$ & $S_\mathrm{peak}$ & $\sigma_\mathrm{RMS}$ & $b_\mathrm{maj}$ & $b_\mathrm{min}$ & PA\\
	& [M$\lambda$]& [mJy] & [mJy/beam]	& [mJy/beam]	& [mas]	& [mas]	& [\degr]\\
	\hline
	natural & - & $624\pm94$ & $414\pm62$& $0.99$& $0.407$& $0.075$& $-10.1$\\
	natural (tapered) & 700 & $612\pm92$  & $412\pm62$ & $1.66$  & $0.587$  & $0.324$  & $-24.75$  \\
	uniform & - & $565\pm 85 $ & $411\pm 62$  & $1.21$  & $0.354$ & $0.057$ & $-9.3$ \\
	\hline		
	\end{tabular}
\end{table*}
% % % % % % % % % % % % % % % % % % % % % % % % %
\section{Results}
\label{sec:Results}
\begin{figure}[!h]
 \centering
 \includegraphics[width=\linewidth]{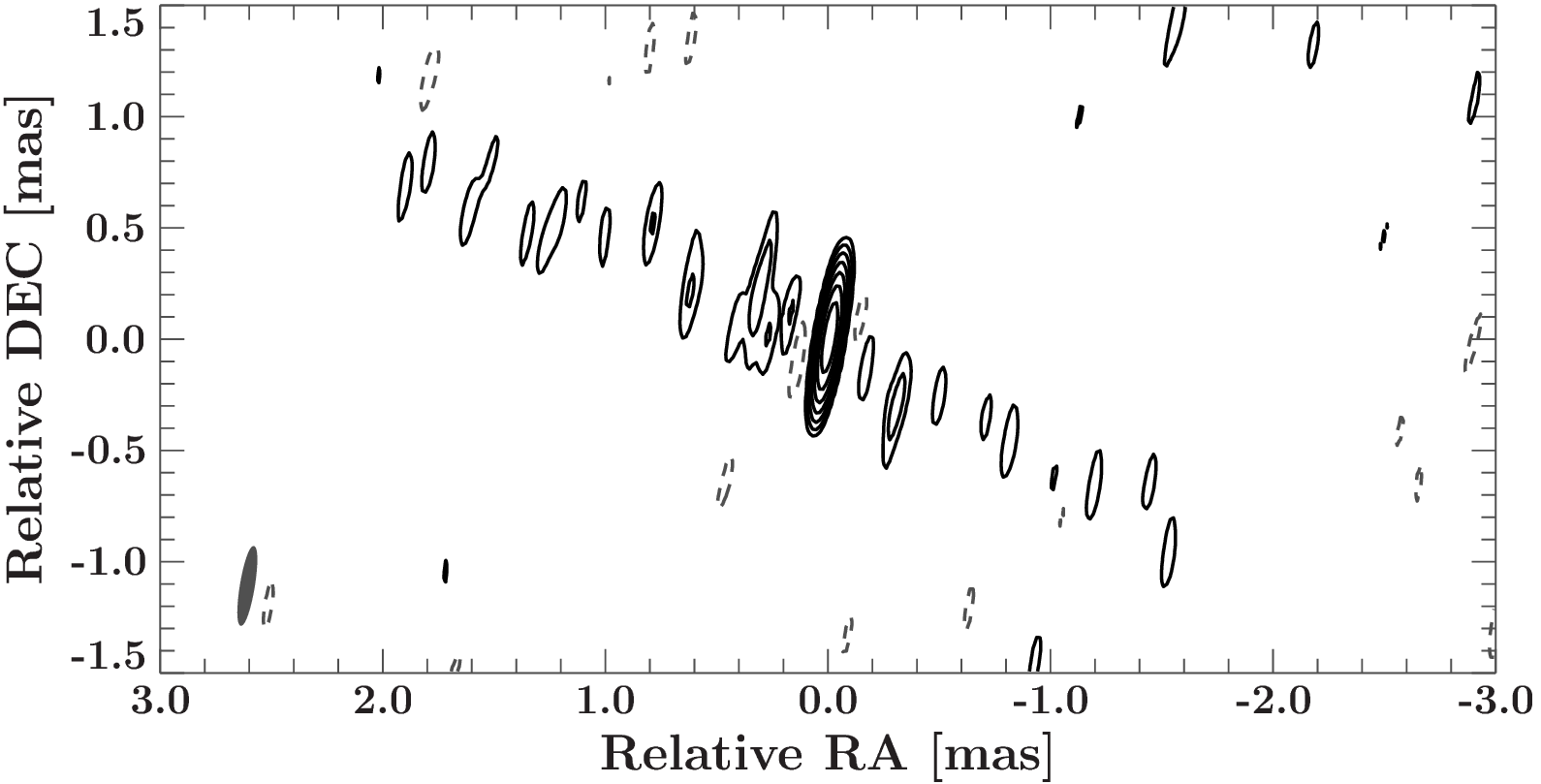}
 \caption{Uniformly weighted GMVA image at 86\,GHz of the innermost structure of the bipolar jets in \object{NGC\,1052}. Contour lines begin at three times the noise level, and increase logarithmically by factors of 2. Image parameters are given in Table \ref{tab:ImageParameters}.}
 \label{fig:unimap}
\end{figure}

\begin{figure}[!h]
 \centering
 \includegraphics[width=\linewidth]{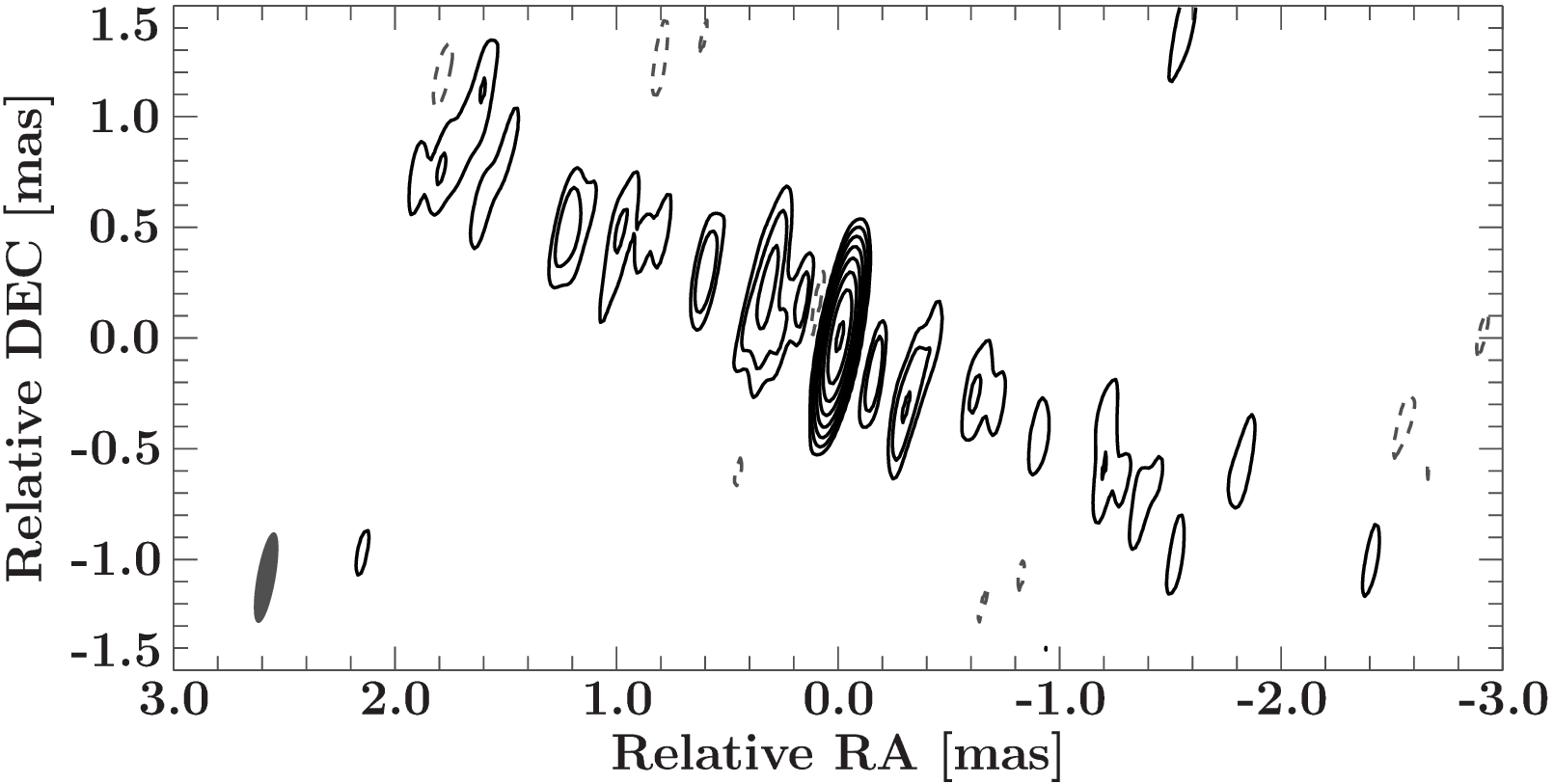}\\
 \includegraphics[width=1\linewidth]{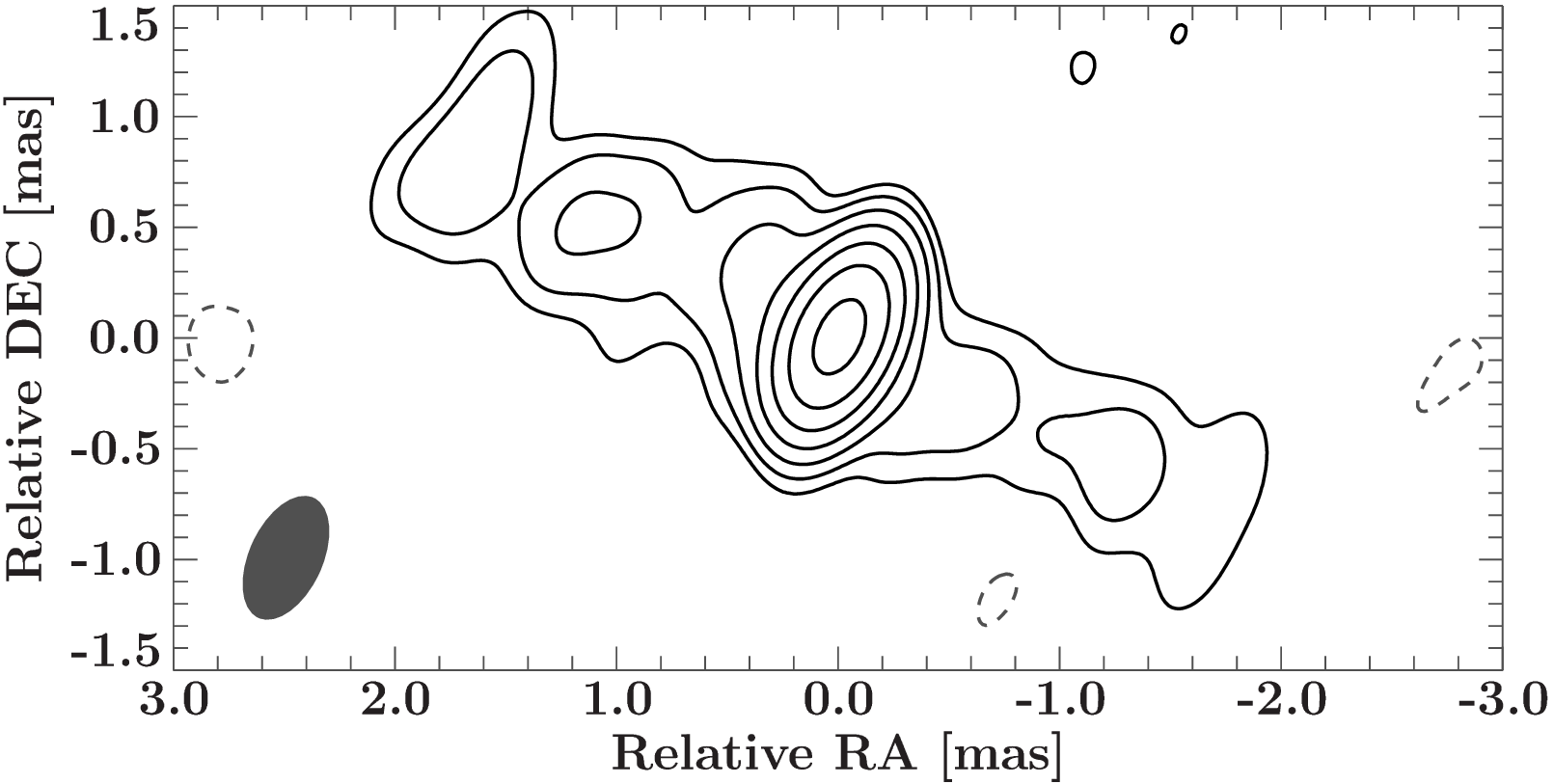}
 \caption{
 Top: naturally weighted GMVA image at 86\,GHz of the innermost structure of the bipolar jets in \object{NGC\,1052}; Bottom: with taper at 700M$\lambda$. Contour lines begin at three times the noise level, and increase logarithmically by factors of 2. Image parameters are given in Table \ref{tab:ImageParameters}.}
 \label{fig:natmap}
\end{figure}

% % % % % % % % % % % % % % % % % % % % % % % % %
The twin-jet system of \object{NGC\,1052} at the highest-angular resolution observed so far for this source is shown in 
Figures~\ref{fig:unimap}
and \ref{fig:natmap} for uniform and natural weighting respectively. It represents the highest-linear-resolution image of an AGN ever taken. The minor axis of the Gaussian image restoring beam, which is oriented close to the east-west jet axis, is only 57\,$\mu$as in diameter, which corresponds to $0.006\,$pc or $7$ light days.
The image reveals a remarkably symmetric structure with one central high-brightness-temperature feature and two fainter jets emanating from there to the east and west. A tapered map is shown in Figure~\ref{fig:natmap}, giving a better impression on the overall structure. The two jets are partially resolved by the interferometric array and show a shallow brightness gradient out to about 2\,mas from the center. Due to the lack of short baselines, some diffuse flux may have been lost. More than 70\,\% of the total correlated flux density is contained inside the central feature, which can be identified with component B/B2 studied in earlier works \citep{Kam01,Kad04a}. Lower-frequency studies suggested the black hole to be located $\sim (0.25$ to $0.50)$\,mas eastward of this region \citep{Kad04a}, thus ascribing the component to the western jet, which was assumed to be the counter jet \citep{Kam01,Ver03,Kad04a}. The striking symmetry in our new higher-frequency and higher-resolution image strongly suggests the true center of activity to be located 
inside the central feature, which then represents the unresolved emission of both jet cores and pinpoints the position of the supermassive black hole. We refer to the unresolved central feature as the \textsl{3\,mm core of \object{NGC\,1052}}.
This is one of the most precisely known locations of an extragalactic supermassive black hole in the Universe. Estimates have been obtained e.g., by \cite{Had11} for \object{M87}, by \cite{Fro13} for \object{CTA\,102}, and \cite{Mue14} for \object{Centaurus\,A}, though these jets are seen at a smaller angle to the line of sight which leads to projection effects.


To further study the flux density distribution, the structure is divided into three parts, i.e., a center, a north-eastern (NE) and a south-western (SW) jet region. We summed up the \textsc{clean}-components within each region based on the naturally weighted image to gain a total flux density. All \textsc{clean}-components within $0.1$\,mas from the origin were attributed to the center. This yields a flux density for the NE and SW jet of $(118\pm18)\,$mJy and $(84\pm13)\,$mJy, respectively. These components were well localised and separated from other jet components. The center contains $(415\pm62)\,$mJy ($\sim67\,\%$) of the total flux density in the image. We determine the position angle $\phi$ of both jets from the distribution of \textsc{CLEAN} components, yielding a weighted mean of $\langle\phi_\mathrm{NE}\rangle = (64\pm 12)\degr$ and $\langle\phi_\mathrm{SW}\rangle = (-120 \pm 13)\degr$ for the NE and SW-region, respectively, as measured from the North, with the errors representing the standard deviation. The \textsc{clean}-components were weighted with their flux density. These values are consistent with the median of $\phi$, i.e. $\phi_\mathrm{NE,median}=66\degr$ and $\phi_\mathrm{SW,median}=-118\degr$. These values are consistent with previous works  \citep{Jon84,Kad04b}.

Our image marks the first time the two jets of \object{NGC\,1052} are detected by VLBI observations at 86\,GHz. A previous 86\,GHz snap-shot observation obtained by  \citep{Lee08} with the Coordinated mm-VLBI Array in 2002 depicted only a single bright and compact feature of \object{NGC\,1052}, most likely due to lack of sensitivity to detect the jet. Our full-time GMVA observation achieved a dynamic range of 650:1, which is remarkable for such a low-luminosity and low-declination source.

The 3\,mm core appears unresolved in the transverse (north-south) direction (i.e., $\lesssim150$\,$\mu$as, which corresponds to $\lesssim1000\,R_\mathrm{S}$ for a black hole mass of $10^{8.2}\mathrm{M_\odot}$).
The width of both jets can be measured, however, at a distance of about $0.2$\,mas east and west of the nucleus. We find a width of $\sim 0.2$\,mas in both cases by fitting two Gaussian components at this distance. Thus, the opening angle of both jets has to be larger than $60^\circ$ within the inner 0.2\,mas. This result is similar to the situation in the inner milliarcsecond of the \object{M\,87} core, which has been interpreted as a sign of a magnetohydrodynamically launched but still de-collimated jet \citep{Jun99}. In a region of such rapid expansion, possibly accompanied by differential expansion  \citep{Vla04}, magnetic field energy can be converted efficiently into bulk kinetic energy, effectively accelerating the jets in this region. This result is in agreement with the abrupt change in brightness that we observe between the 86\,GHz (3\,mm) core and \object{NGC\,1052}'s jets.

% % % % % % % % % % % % % % % % % % % % % % % % %
\section{Discussion}
\label{sec:Discussion}
% % % % % % % % % % % % % % % % % % % % % % % % %

\subsection{Orientation of the twin-jet system}
\label{sec:Discussion:Jet-Orientation}
\begin{figure}
	\includegraphics[width=1\linewidth]{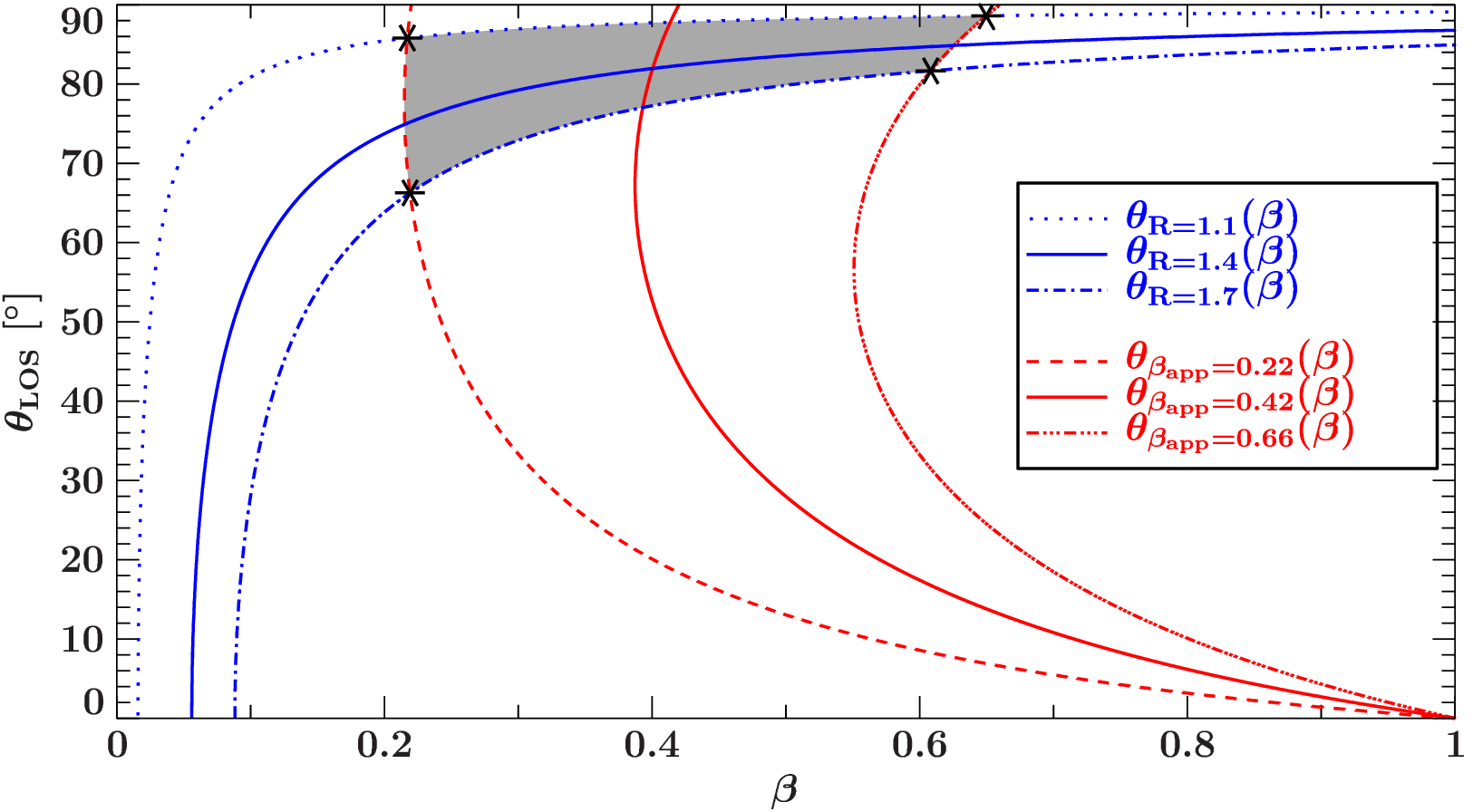}
	\caption{The angle of the jet to the line of sight $\theta_\mathrm{LOS}$ depending on the jet velocity $\beta$ constrained by the jet-to-counter jet ratio equation~\ref{eq:LOS}) measured at 86\,GHz (blue) and the apparent jet velocities (equation~\ref{eq:Beta})	at 43\,GHz, giving the mean, minimum and maximum value (red). The allowed parameter space for $\theta_\mathrm{LOS}$ and $\beta$ is highlighted by the grey-shaded region.}
	\label{fig:NGC1052LOS}
\end{figure}

% % % % % % % % % % % % % % % % % % % % % % % % %
To determine the orientation of the twin-jet system with respect to the line of sight, we measured the flux density in the CLEAN model of our final image for the two jets separately, excluding the central feature from the regions used for the flux-density measurements of both jets (see Sect.~\ref{sec:Results}). We thus found a flux-density ratio of $R=S_\mathrm{east}/S_\mathrm{west} = 1.4 \pm 0.3$ and calculate the angle of the jet to the line of sight $\theta_\mathrm{jet}$, which is linked to $R$ for a continuous jet via
\begin{equation}
\frac{S_\mathrm{east}}{S_\mathrm{west}} = \left(\frac{1+\beta\cos\theta_\mathrm{LOS}}{1-\beta\cos\theta_\mathrm{LOS}}\right)^{2-\alpha},
    \label{eq:LOS}
\end{equation}
where $\beta$ is the intrinsic jet velocity and $\alpha$ the spectral index. Here, we use the measured spectral index of the jet $\alpha_\mathrm{jet}\approx -1$ from lower frequencies \citep{Kad04a}. The inclination angle is also related to the intrinsic and apparent jet speeds $\beta$ and $\beta_\mathrm{app}$ via
\begin{equation}
\beta_\mathrm{app} = \frac{\beta\sin\theta_\mathrm{LOS}}{1-\beta\cos\theta_\mathrm{LOS}} \quad .
\label{eq:Beta}
\end{equation}
Since $\beta_\mathrm{app}$ has not been measured at 86\,GHz for \object{NGC\,1052}, we adopt a range of $\beta_\mathrm{app,min} = 0.22$ to $\beta_\mathrm{app,max} = 0.66$ from a kinematic analysis at 43\,GHz over four years of VLBA observations (A.-K. Baczko et al., in prep, see also \citealt{Bac15} and \citealt{Bac15b}). This assumes a conservative range of jet speeds, that includes values from previous studies at smaller frequencies, as well \citep{Ver03,Lis13}. However, \cite{Boeck12} reported a maximum velocity of $\beta_\mathrm{app}=0.42$, that is close to the averaged speed of $\beta_\mathrm{app}=0.46\pm 0.05$ derived from the coeval 43\,GHz kinematic study (A.-K. Baczko et al., in prep).
The allowed range of parameters is highlighted by the grey-shaded area in Fig.~\ref{fig:NGC1052LOS}. This yields $64^\circ \lesssim \theta_\mathrm{LOS} \lesssim 87^\circ$ and $0.21\lesssim\beta\lesssim 0.64$. The range of $\theta_\mathrm{LOS}$-values shows the western jet of \object{NGC\,1052} to be oriented closer to the line of sight at 86\,GHz than at lower frequencies.

The orientation of the twin-jet system in \object{NGC\,1052} has previously been estimated using cm-wavelength multi-frequency and multi-epoch observations to lie in the range $67^\circ$ to $72^\circ$, i.e., with the eastern jet being the approaching one and the western jet receding. This discrepancy might be explained by jet curvature on the innermost scales or by a slight intrinsic asymmetry  between the two jets.
In fact, the overall structure in our 86\,GHz GMVA image is similar to that shown at 43\,GHz in July 1997 \citep{Ver03} but substantially more symmetric than in the 43\,GHz image of December 1998 \citep{Kad04a}. In fact, the substantial difference in the brightness ratio of the two jets between these two 43\,GHz VLBA observations indicates a possible intermittent and asymmetric injection of new plasma at the bases of both jets, which is also seen in a series of 43\,GHz VLBA observations in 2005/2006. We will investigate this further in a later publication.
This, along with a possible under-estimation of the strength of free-free absorption of the components at distances between 3\,mas and 5\,mas down the two jets \citep{Kad04a} might have contributed to an under-estimation of the inclination angle of the twin-jet system. 
All measurements of the inclination angle agree in the sense that the twin-jet system must be oriented close to the plane of the sky. As asymmetries are observed at lower frequencies, we cannot claim stability at 86\,GHz, too. Our results represent the first snapshot observation at this frequency for \object{NGC\,1052}, revealing the double-sided jet system. Therefore future observations at the same frequency are needed to give estimates on the stability of the jet system at 86\,GHz.

% % % % % % % % % % % % % % % % % % % % % % % % %
\subsection{The core region -- Modeling the central emission feature}
\label{sec:Discussion:Core-Region}
% % % % % % % % % % % % % % % % % % % % % % % % %
To estimate the size of the emission region we fit the central feature in the uniformly-weighted 86\,GHz GMVA image with an elliptical Gaussian component by replacing all \textsc{clean}-components within 0.05\,mas of the image center. The fit was performed with \textsc{ISIS} \citep{Hou00} with a direct connection to \textsc{Difmap} \citep{She94}. This allowed us to determine statistical errors on the fit parameter based on $\chi^2$-statistics \citep{Gro14}. Table~\ref{tab:ModelfitParameters} lists the resulting fit parameters with the statistical error given at the 1$\sigma$ level. In the uniformly-weighted image we tested whether the component is unresolved, based on the signal-to-noise ratio. We calculate the resolution limit $\theta_{\mathrm{lim},\psi}$ as 
\begin{equation}
 \theta_{\mathrm{lim},\psi}=2^{2-(\beta/2)}\left[ \frac{\ln 2}{\pi}\ln\left(\frac{\mathrm{SNR}}{\mathrm{SNR}-1}\right)\right]^{1/2}b_\psi,
 \label{equ:thetalim}
\end{equation}
where $\beta=0$, due to uniform weighting, $\mathrm{SNR}=187$ is the ratio of the peak flux density and the noise level at the position of the component and $b_\psi$ is the beam axis at the position angle $\psi$ of the component \citep{Lob05}. The calculated limits for the minor and major axis of the component are larger than the fit, thus the elliptical Gaussian component is unresolved. The limit leads to an upper limit on the size of the emitting region along the jets of $<8\,\mu$as. A widely used, empirical resolution limit is given by one fifth of the beam size which yields an upper limit on the size of $<12\,\mu$as. This estimate is slightly larger than that derived using Eq. \ref{equ:thetalim}. 
We applied a third approach for deriving the size of the central region by fitting two point sources in the direction of the jets. From an initial separation of $0.1\,\mu$as we enlarged their distance until major changes in the residuals were observed. This was the case for a component separation of$\sim30\,\mu$as, which is close to half the beam size. Thus using half the beam size would give the most conservative upper limit on the size of the emission region. However, the central feature is clearly unresolved and having the power of super-resolution in VLBI observations  \citep{Mar12b} in mind, a smaller size is likely for our 3\,mm observation.
All derived values are listed in Table~\ref{tab:ModelfitParameters}. 

The higher sensitivity that is to be expected from near-future VLBI arrays may improve the over-resolution power in the observations far above the diffraction limit currently achieved  \citep{Mar12b}, thus leading to an even better modeling of the central region of \object{NGC\,1052}.

\begin{table*}
	\centering
	\caption[]{Best core model-fit parameters for uniform weighting. Uncertainties of the component parameters are given at the 1$\sigma$ level for one interesting parameter.}
	\label{tab:ModelfitParameters}
	\begin{tabular}{@{}ccccccc@{}}
	\hline\hline
	Method & $S_\mathrm{comp}$  &  $a_\mathrm{maj}$ & $a_\mathrm{min}$ & PA & $T_\mathrm{b}$  \\
	 & $[$mJy$]$  & [$\mu$as]     & [$\mu$as]	   & [$^\circ$]& [K] \\
	\hline
	Elliptic Modelfit & $409^{+3}_{-2\,\mathrm{stat}}\pm62 _\mathrm{sys}$ & $14^{+25}_{-14}$&0 &$-2.5$ & --\\
	Circular Modelfit & $409^{+2}_{-2\,\mathrm{stat}}\pm62 _\mathrm{sys}$&$\le 5$& -- &-- & $\geq3\times 10^{12}$\\
	Lobanov (2005) & $409^{+3}_{-2\,\mathrm{stat}}\pm62 _\mathrm{sys}$ & $\le39$ & $\le8$ & $-2.5$ & $\geq 2\times 10^{11}$ \\
	1/5th beam & $409^{+3}_{-2\,\mathrm{stat}}\pm62 _\mathrm{sys}$ & $\leq57$ & $\leq12$ & $-2.5$ & $\geq1\times 10^{11}$\\
	1/2th beam & $409^{+3}_{-2\,\mathrm{stat}}\pm62 _\mathrm{sys}$ & $\leq88$ & $\leq39$ & $-2.5$ & $\geq1\times 10^{11}$\\
	\hline		
\end{tabular}
\end{table*}

\subsection{Magnetic field estimate}
\label{sec:Discussion:Magnetic-Field}
% % % % % % % % % % % % % % % % % % % % % % % % %
\subsubsection{Radio Emission Mechanisms and Location}
High-energy electrons in a strong magnetic field near the base of an AGN jet are accelerated and emit radio synchrotron radiation. The emitting electrons lose their energy quickly as they radiate. If the emitting particles change direction out of our line of sight or are not re-accelerated efficiently enough, the jet is no further visible at this frequency beyond the point at which those particles have lost most of their energy. The 86\,GHz image of the jets in \object{NGC\,1052} shows a steep brightness drop from the core brightness to the much lower jet brightness within the first 15$\,\mu$as or less. 

It is also possible that the central emission could originate at least in part from the accretion disk. Models for accretion disks indicate that the emission can reach similar brightness temperatures as in our case, but these models are strongly parameter dependent  \citep{Asa09}. A detailed investigation of possible accretion disk models for \object{NGC\,1052} is beyond the scope of this paper. Hence, we consider here the case with the fewest assumptions, i.e., that the flux density of the central region stems completely from the jet. Interpreting the central feature in this way makes it possible to derive the magnetic field in this area via synchrotron losses, assuming that the 86\,GHz emission disappears owing to synchrotron cooling.
\vspace{0.2cm}

\subsubsection{Estimating the magnetic field inside the 86\,GHz and 22\,GHz cores}
The magnetic field strength in the emitting region can be estimated with the following line of argument \citep{Ryb04,Boe12}.
The synchrotron cooling time of a particle radiating at 86\,GHz is given by Larmor's formula for relativistic particles. Assuming a random pitch angle distribution, the emitted power for an electron is
\begin{equation}
 P=\frac{4}{3} c\,\sigma_T\, u_B\,\beta^2\,\gamma^2\,,
 \label{equ:power_sync}
\end{equation}
being $\gamma$ the Lorentz factor and $\beta$ the ratio of the speed of the radiating electrons to the speed of light, $\sigma_T$ the Thomson cross section, and $u_B$ the energy density of the magnetic field.

We can express the energy loss of the electron due to radiation in terms of the change of its Lorentz factor
\begin{equation}
\left(\frac{\mathrm{d}\gamma}{\mathrm{d}t}\right) =
-\frac{4}{3} \sigma_T \frac{u_B}{m_e\cdot c}\,\gamma^2\beta^2\;,
\label{equ:enloss}
\end{equation}
with $m_e$ the electron mass.
Integration yields the cooling time, dependent on the energy and radiation rate of the electron, for a given Lorentz factor 
\begin{equation}
 t_\mathrm{c}=\frac{3}{4}\frac{m_e\,c}{\gamma \, \sigma_T\,u_B\,\beta^2}\;.
\end{equation}
Inserting the energy density of the magnetic field and the critical frequency (frequency at which the spectral output peaks),
\begin{equation}
 \nu_\mathrm{c}\equiv\frac{3}{4\pi}\frac{e\,B}{m_e\,c}\gamma^2,
\end{equation}
with $e$ the charge of the electron and $B$ the magnetic field, the cooling time becomes
\begin{equation}
t_\mathrm{c}=3\sqrt{3\pi}\sqrt{\frac{m_e\,c\,e}{\nu_c}}\frac{1}{\sigma_T\,\beta^2}B^{-3/2} = 5.4\times10^6\, B^{-3/2} /[\mathrm{G^{-2/3}}]\,[\mathrm{s}]\;,
\end{equation}
Inserting the Thomson cross section $\sigma_T=6.65\times\,10^{-25} \mathrm{cm}^2$, the electric charge $e=4.8\times 10^{-10} \mathrm{statC}$, the mass of the electron $m_\mathrm{e}=9.11\times 10^{-28}\mathrm{g}$ and a critical frequency of $\nu_c\simeq 86\;\mathrm{GHz}$.

As the emitting particles are advected with the flow, this expression can be related to the size of the core by means of the bulk speed of the jet.

Setting the travel distance of the electrons during this time equal to the assumed resolution limit, we obtain for the case of synchrotron cooling
\begin{equation} \label{cool}
 B_{\mathrm{\,sc,\,}d}=\left( \frac{d/[\mathrm{cm}]}{\beta\,\mathrm{[\mathrm{cm\,s^{-1}}]} \times\,5.4\,\times 10^6\,\mathrm{s}} \right)^{-2/3}\, [\mathrm{G}],
\end{equation}
which depends on the bulk speed $\beta$ of the jet and the size $d$ of the emitting region. This gives the mean magnetic field intensity of the region from which the radiation is emitted.

\begin{figure}
 \centering
 \includegraphics[width=0.95\linewidth]{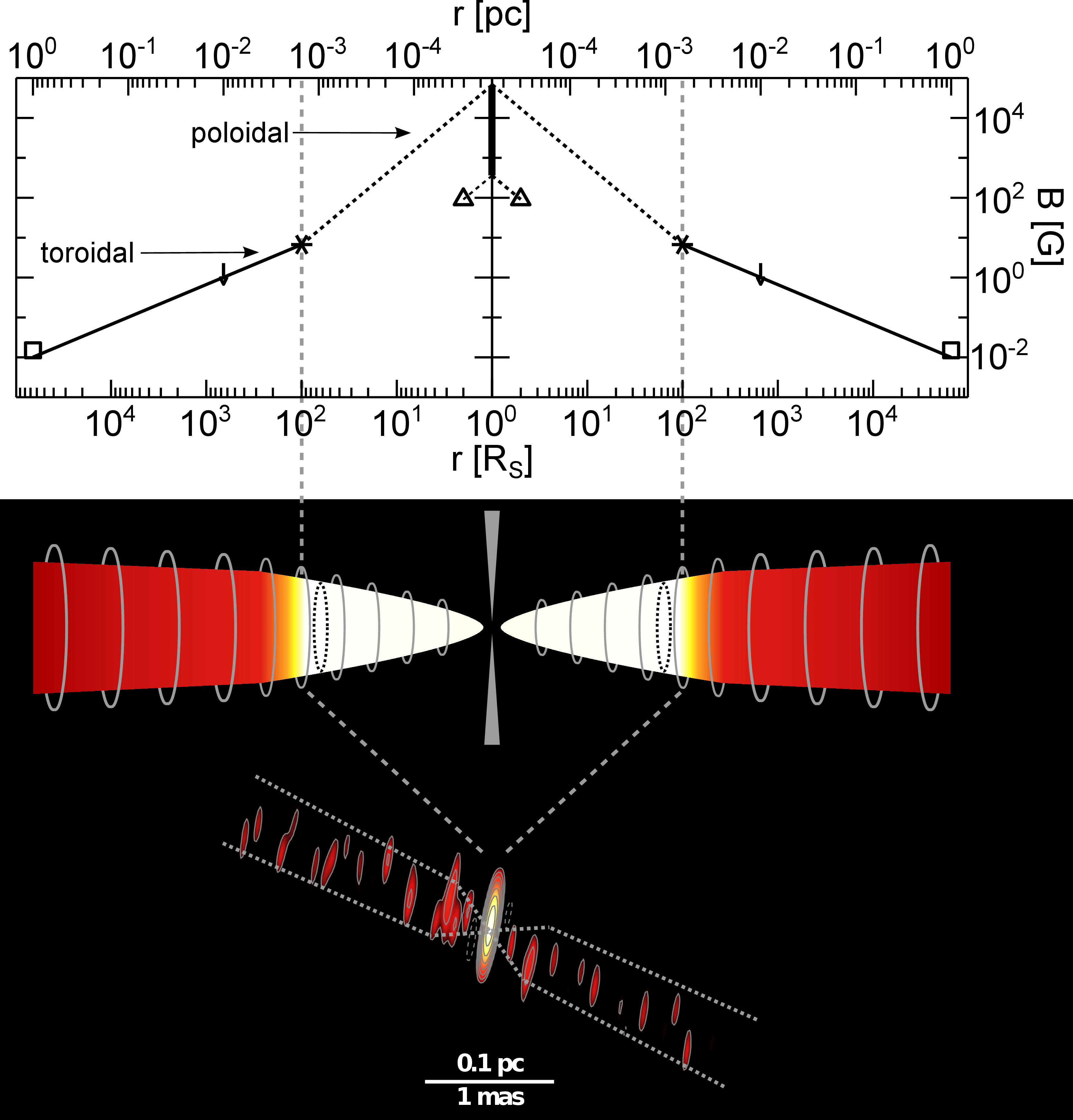}
 \caption{Sketch of the jet-launching region. The mm-VLBI image (bottom) resolves structures larger than $\sim 100$ to $400$ Schwarzschild radii ($R_\mathrm{S}$)around the central supermassive black hole, that is assumed to be maximally rotating with $a=1$. In the innermost region, both jets have extremely large opening angles of $>60^\circ$. The regions at which the two jets become optically thick, usually referred to as the two jet cores (dashed ellipses in middle panel), are located at even smaller scales ($<100 R_\mathrm{S}$). The magnetic field (top panel) has a toroidal configuration $B\propto r^{-1}$ (solid line). Near the black hole, a poloidal magnetic field may start to dominate (dashed lines). The asterisk and triangle show the measured magnetic field from synchrotron cooling using the most conservative and the lowest possible estimate of the size of the central region, respectively.  From the former the toroidal field out to 1pc and the poloidal field down to $1\,R_\mathrm{S}$ are extrapolated, while from the latter only the poloidal field down to $1\,R_\mathrm{S}$ is extrapolated. This provides a strength of the magnetic field near the event horizon at $\sim1\,R_\mathrm{S}$ of at least $360\,$G reaching values up to $6.9\times10^4\,$G. The thick line in the middle of the upper panel indicate the allowed range for the magnetic field at $1\,R_\mathrm{S}$. The down-oriented arrow and the square indicate the upper limit from SSA and core-shift measurements, respectively.}
\label{fig:sketch}
\end{figure}

Therefore, the smallest magnetic-field estimate follows from the most conservative resolution limit, which leads to a size along the jet of $\lesssim30\,\mu$as ($\sim 0.003\,$pc or $\sim 200\,R_\mathrm{S}$). Interpreting this region as blended emission from both jet bases, this size constrains the distance of the central engine and the base of the 3\,mm core to be $\lesssim100\,R_\mathrm{S}$ (see Fig.~\ref{fig:sketch}). The largest estimate of the magnetic-field strength results from the smallest possible size of the emission region of $4\,R_\mathrm{S}$ ($0.6\,\mu$as or $0.06\,$mpc), which is likely to be an underestimation.

As discussed in section 4.1. a kinematic study at 43\,GHz revealed a range of $0.22 \leq \beta_\mathrm{app}\leq0.66$, resulting in an averaged bulk speed of $\beta_\mathrm{app}=0.46\pm 0.05$ for the overall jet flow (see also \citealt{Bac15} and \citealt{Bac15b}). Both, the bulk speed and the cooling distance have the same weight in deriving $B_\mathrm{sc,d}$ (compare Eq.~\ref{cool}). However, the large uncertainity in the cooling distance makes its impact in Eq.~\ref{cool} larger. Therefore, the following discussion will first focus on the different estimates on the size of the central region and assumes the averaged speed resulting from the 43\,GHz analysis. Taking the projection effects into account (compare Sect.~\ref{sec:Discussion:Jet-Orientation}), the changes are negligible in the scope of the herein assumed accuracy, resulting in $\beta\simeq 0.5$ for this case. After deriving the magnetic field at different regions of the jets, we will shortly discuss the impact of the full range of velocities (from the minimal to the maximal measured ones) on our results.

Assuming a bulk speed of $\beta\simeq 0.5$ and the cooling distance ranging from $0.3\,\mu$as to $15\,\mu$as, Eq.~\ref{cool} provides a range of $6.7\,$G to $91\,$G for the magnetic field at the edge the central region.

Another independent approach gives an upper limit on the magnetic field at 0.01\,pc (0.1\,mas) distance from the black hole.
It can be derived from the synchrotron self absorption (SSA) spectrum of the central component. The peak frequency of the SSA, assuming equipartition between the magnetic field and the non-thermal electrons is
\begin{equation}
 \nu_\mathrm{m} \sim 3.2\,B^{1/5}/[\mathrm{G^{1/5}}]\,S_m/[\mathrm{Jy}]\,\theta^{-4/5}/[\mathrm{mas^{-4/5}}]\,(1+z)^{1/5}\;[\mathrm{GHz}]\,,
\end{equation}
resulting in units of GHz for $\nu_m$. $\theta$ is the angular size of the source \citep{Kel81}, $z=0.005$ is the redshift and $S_m$ is the maximum value of the flux density. Using the brightness temperature $T_b$ of a given component, the magnetic field strength can be thus derived as
\begin{equation}
 B_\mathrm{\,SSA}\sim 4.57\times 10^{19} \nu_m/[\mathrm{GHz}]\, T_b^{-2}/[\mathrm{K^{-2}}]\, (1+z)\;[\mathrm{G}].
\end{equation}
The spectrum of the central core component peaks at $\simeq 22$\,GHz \citep{Kad04a}. As free-free absorption also takes place, this gives us an upper limit on the SSA peak frequency $\nu_\mathrm{m}$. The innermost component B2b has $T_\mathrm{b} \simeq 2.0\times 10^{10}\,\mathrm{K}$ \citep{Kad04a}. Converting $T_\mathrm{b}$ and $\nu_\mathrm{m}$ to the jet rest-frame (using a viewing angle of $86^\circ$, spectral index $\alpha=0.0$ and flow speed $\beta=0.25$) gives $T_\mathrm{b}^*=2.2\times 10^{10}\,$K and $\nu_\mathrm{m}^*=23\,$GHz. Adopting a core size of $(0.21\pm0.01)$~mas leads to $B_\mathrm{\,SSA}<2.5\,\mathrm{G}$. This value can be thus considered as an upper limit on the magnetic field intensity at a distance of $0.01\,$pc ($0.1$~mas) along the jet.
This upper limit is in agreement with the expectations for a toroidal magnetic field decreasing as $r^{-1}$ between the 3\,mm core and 0.1\,mas ($1$\,G would be expected in this scenario for the largest black hole-3\,mm core distance of $0.0015\,$pc), assuming that the jet is adiabatically expanding. 

A third independent estimate of the magnetic field in the jet is given by the measurement of the core-shift effect \citep{Zam14}, assuming equipartitioning of the energy between the magnetic field and the non-thermal particles. This approach yields a magnetic field of $15\,$mG at 1\,pc ($10\,$mas).
This is in remarkable agreement with the expectation for a $r^{-1}$ decrease of the 6.7\,G field from $0.0015\,$pc out to 1\,pc, for which $10$\,mG would be expected.
Together, these three estimates of the magnetic field along the twin jets of \object{NGC\,1052} yield a self-consistent picture of a toroidal magnetic field dominating outside the 3\,mm core and decreasing outwards proportional to $r^{-1}$. So far, an $r^{-1}$ dependence has been adopted in the literature to estimate the inner magnetic field close to the black hole based on plausibility \citep{Osu09,Pus12}.

The gradient of the toroidal magnetic field is thus well measured on scales from the 3\,mm core to 1\,pc, and may be  extrapolated further inwards as a lower limit. At some distance closer than $0.0015\,$pc, near the central engine, a poloidal magnetic field with a $r^{-2}$ dependence is expected to start dominating over the toroidal field \citep{Kom12}, that is defined by the spin parameter, that we assumed as $a=1$ for the case of a probably rapidly rotating black hole in \object{NGC\,1052} \citep{Bre09}. If effective magnetic acceleration of the jet takes place, converting magnetic into bulk kinetic energy, this gradient might steepen considerably. 

We assume a toroidal magnetic field from the 3\,mm core until $2\,R_\mathrm{S}$, where the poloidal field starts to dominate \citep{Kom12}. This leads to a dependence of the magnetic field at $1\,R_\mathrm{S}$ on the size of the emitting region $d$ of $B_{\mathrm{\,sc,\,1\,R_S}}\propto d^{1/3}$. Assuming an unrealistically small size for this region of $4\,R_\mathrm{S}$ gives the lowest possible magnetic field strength at $1\,R_\mathrm{S}$ of $\simeq 360\,$G. At these scales one should check whether the temporal shift due to the central mass should be taken into account. The scaling factor of 1.15 is negligible in the scope of the uncertainty of the magnetic field.
The maximum upper limit on the emission region that is still in agreement with our analysis is half of the beam. Assuming a poloidal magnetic field from this distance towards the central engine gives an upper limit on the magnetic field at $1\,R\mathrm{_S}$ of $\leq6.9\times 10^4\,$G. An overview of all the magnetic field values derived in the previous paragraphs is given in Table \ref{tab:MagneticField}. 

The range of the magnetic field follows by assuming an averaged bulk speed of $\beta_\mathrm{app}=0.5$. As discussed in Section~\ref{sec:Discussion:Jet-Orientation} it is the highest averaged value until now, but not far away from other estimates. For example assuming a velocity of $\beta_\mathrm{app}=0.4$ \cite[maximum values reported by][]{Lis13,Boeck12} gives only slightly deviations, leading to $310\,\mathrm{G}\leq B_\mathrm{sc,\,1\,R_\mathrm{S}} \leq 6.0\times 10^4\,\mathrm{G}$. Magnetic fields with boundaries from $200\,$G to $430\,$G, and $4.0\times10^4\,$G to $8.3\times10^4\,$G are implied, following the highest and lowest velocity values measured at 43\,GHz (A.-K. Baczko et al., in prep, see also \citealt{Bac15} -- compare Fig.~\ref{fig:NGC1052LOS}). Therefore, the impact of the uncertainity in the apparent bulk speed on the magnetic field strength is negligible compared to that resulting from the size of the central region. This is probably caused by the larger uncertainty in the region size compared to the velocity estimates on the jet flow.

An important target with which to compare our results is \object{M\,87}. It is located at a distance of $16.7\,$Mpc \citep{Jor05} similar to \object{NGC\,1052}. The mass of the central black hole in \object{M\,87} ($6\times 10^9 \,\mathrm{M_\odot}$ \citet{Geb09}) is about two orders of magnitude larger than in \object{NGC\,1052}, hence, the Schwarzschild radius of the central black hole in \object{M\,87} is about two orders of magnitude larger than that of \object{NGC\,1052}. This is, however, partially countered by the difference in inclination angle to the line of sight. At $15^\circ$ to $25^\circ$ inclination angle  \citep{Acc09,Bir99}, all linear scales in the \object{M\,87} jet are affected by a projection with a factor of $0.26$ to $0.42$. This results in $\sim7\mu$as corresponding to $1\,R_\mathrm{S}$ in \object{M\,87}.
Thus, formally, VLBI observations can probe about 30 times smaller structures in units of Schwarzschild radii in \object{M\,87} than in \object{NGC\,1052}. Two additional effects play in favor of \object{NGC\,1052}: \textsc{i)} distances along the two-jet system can be measured with double precision, if intrinsic symmetry between the two jets is assumed, and \textsc{ii)} the orientation of the projected jet system is nearly perfect in east-west direction and is conveniently well aligned with the minimal width of the elliptical restoring beam for Northern VLBI arrays to gives us the best resolution along the jet axis.

Based on core-shift measurements, the magnetic field inside the 43\,GHz jet core of \object{M\,87} has been measured \citep{Kin14} to lie in the range $(1$ to $15)$\,G (at a projected distance of $\sim 6\,R_\mathrm{S}$ to the jet base). Based on their analysis \cite{Kin14} excluded magnetic fields as strong as $(10^3$ to  $10^4)$\,G. These are usually assumed to explain the formation of AGN jets via extraction of magnetic energy from the rotational energy of the black hole via the \cite{Bla77} mechanism, and which has been estimated for blazars \citep{Osu09,Pus12}.
It was shown that the jet base in \object{M\,87} is located within $(14\;\mathrm{to}\;23)\,R_\mathrm{S}$ of its radio core at 43\,GHz \citep{Had11}. At frequencies of 228\,GHz, corresponding to 1.3\,mm wavelength, the size of the \object{M\,87} jet core has been measured to be as small as $5.5\,R_\mathrm{S}$. This small size is comparable to the  predicted inner edge of a co-rotating accretion disk around a Kerr black hole \citep{Doe12}.
Based on this observation \cite{kin15} derived the magnetic field inside the 228\,GHz-core to $58\,\mathrm{G}\leq B_\mathrm{tot}\leq127\,\mathrm{G}$ for a region size of $21\,\mu\mathrm{as}\leq \theta \leq 25.5\,\mu\mathrm{as}$. Extrapolating the result from core shift measurements \citep{Had11} down to this frequency, the 228\,GHz-core is located at a deprojected distance of $(2.5$ to $4)\,R_\mathrm{S}$ \citep{Doe12}. Therefore the magnetic field estimate is in agreement with the lower edge of our magnetic field derivation inside the 86GHz-core ($\sim90\,$G for a distance from the black hole of $2\,R_\mathrm{S}$). 
The jet shows a wide opening angle of $\sim 60^\circ$ in the innermost region and is collimated within $\sim 30$ to $100\,R_\mathrm{S}$ of the core \citep{Jun99}. This provides circumstantial evidence that the 1.3\,mm jet core is located very close to the black hole. This hypothesis cannot be confirmed by approaching the same location from the counter-jet side as the flux density of the counter jet \citep{Kov07} in \object{M\,87} is too low and the angle to the line of sight \citep{Acc09} is only ($15$ to $25$)$^\circ$.
An additional offset between the black hole and the jet base, as well as a steep magnetic-field gradient might still increase the peak magnetic-field strength near the black hole into the Blandford \& Znajek regime, but this hypothesis is difficult to test in more inclined sources with a high jet-to-counter-jet brightness ratio like that in \object{M\,87}. An even higher magnetic field strength on scales of 0.01\,pc has been reported by \cite{Mar15} for \object{PKS\,1830-211} , leading to tens of gauss on these scales.

Summarizing, we have obtained independent estimates of the magnetic field intensity at two different distances from the black hole ($0.5$~mpc and $10$~mpc) using basic synchrotron theory, which, together with a third estimate from core-shift at 1~pc  \citep{Zam14}, are consistent with the magnetic field intensity falling with $r^{-1}$, i.e., implying that the magnetic field structure is dominated by the toroidal component already at sub-parsec scales. The peak magnetic field in both scenarios exceeds the value of $\sim 200$\,G needed to launch the jet with kinetic power of $5 \times 10^{41}$erg\,s$^{-1}$ (as derived by \citet{Kad03}) via the Blandford \& Znajek mechanism \citep{Bla77,Kom12}.

\begin{table*}
 \centering
 \caption[]{Magnetic field values.}
 \label{tab:MagneticField}
 \begin{tabular}{@{}ccccccc@{}}
  \hline\hline
	Method & region size $d$ & region size $d$ &$B_{\mathrm{sc\,}d}$ & $B_{\mathrm{sc,\,1\,pc}}$ & $B_{\mathrm{sc,}\,1\,R_\mathrm{s}}\tablefootmark{a}$ &$B_{\mathrm{sc,}\,1\,R_\mathrm{s}}$\tablefootmark{b} \\
	& $[\mu\mathrm{as}]$&$[$mpc$]$ & $[$G$]$ & $[$G$]$ & $[$G$]$ & $[$G$]$ \\\hline
	$4\,R_\mathrm{S}$ & $0.6$ & $0.06$&$\geq91$ & $\geq0.003$ & $\simeq360$ & $\simeq360$\\
	Lobanov (2005) & $8$ &$0.8$& $\geq16$ & $\geq0.007$ & $\geq880$ & $\leq1.3\times 10^4$ \\
	1/5th beam\tablefootmark{c} & $12$ & $1.2$ &$\geq12$ & $\geq0.008$ & $\geq1000$ & $ \leq2.1\times 10^4$\\
	half a beam\tablefootmark{c} & $30$ & $3.1$ &$\geq6.7$ & $\geq0.01$ & $\geq1400$ & $\leq6.9\times 10^4$\\
	\hline		
 \end{tabular}
 \tablefoot{
 \tablefoottext{a}{assuming $B\propto r^{-1}$ for $r>2\,R_\mathrm{S}$ and $B\propto r^{-2}$ for $r<2\,R_\mathrm{S}$;}
 \tablefoottext{b}{ $B\propto r^{-2}$ for $r<d$;}
 \tablefoottext{c}{ beam at 86\,GHz, value used is FWHM of the beam at Position angle 64$^\circ$ along the jet axis.}
 }
\end{table*}

% % % % % % % % % % % % % % % % % % % % % % % % %
\section{Conclusion \& Summary}
\label{sec:Conclusion}
% % % % % % % % % % % % % % % % % % % % % % % % %

Our analysis of the highest-resolution VLBI observation of \object{NGC\,1052} at 86\,GHz with the GMVA from 2004-10-09 shows the twin-jet of \object{NGC\,1052} extending from a single core without any indication of absorption due to the obscuring torus seen at lower frequencies. 
The average flux density ratio between both jets is determined to be $1.40\pm0.30$ (being the eastern jet brighter). In combination with measurements of the apparent jet velocity at lower frequencies, we determine the angle of the jet to the line of sight to be $\sim 86\degr$. Instead of seeing two separate radio cores, the two jets are extending from a single central region with a size of $<200R_\mathrm{S}$. The center in our image of \object{NGC\,1052} may include the radio core of each jet, contributing to the emission in this part. From an estimate of the synchrotron cooling time it was possible to derive a magnetic field at $1\,R_\mathrm{S}$ of $360\,\mathrm{G}<B<6.9\times 10^4\,$G.
We have shown that it is feasible to reach sufficient sensitivity with VLBI at 86\,GHz to detect the two jets. A kinematic study and higher sensitivity observations at 86\,GHz are crucial to further constrain the jet geometry and compare the evolution of the jet with numerical models of bipolar jet production especially in terms of asymmetries in the jet outflow.
Adding space-based VLBI will improve the resolution in the north-south direction, and going to even higher frequencies will improve the overall resolution. This makes \object{NGC\,1052} one of the very few, very close and bright targets to test the unification scheme of AGN and jet formation on scales of several light-days in a twin-jet system.

\begin{acknowledgements}
We thank Alan Roy for his insightful comments, which improved the manuscript. 
We acknowledge support and partial funding by the Deutsche Forschungsgemeinschaft grant WI 1860-10/1 and GRK 1147, the Spanish MINECO project AYA2012-38491-C02-01, AYA2013-4079-P, and AYA2013-48226-C03-02-P, the Generalitat Valenciana projects PROMETEOII/2014/057, and PROMETEOII/2014/069, as well as by the COST MP0905 action `Black Holes in a Violent Universe'.
This research has made use of the software package \textsc{ISIS}  \citep{Hou00} and a collection of \textsc{ISIS} scripts provided by the Dr. Karl Remeis Observatory, Bamberg, Germany at http://www.sternwarte.uni-erlangen.de/isis/, as well as an \textsc{ISIS}-DIFMAP fitting setup by C. Grossberger \citep{Gro14}.
This research has made use of the NASA/IPAC Extragalactic Database (NED) which is operated by the Jet Propulsion Laboratory, California Institute of Technology, under contract with the National Aeronautics and Space Administration.
This research has made use of data obtained with the Global Millimeter VLBI Array (GMVA), which consists of telescopes operated by the MPIfR, IRAM, Onsala, Mets{\"a}hovi, Yebes and the VLBA. The data were correlated at the correlator of the MPIfR in Bonn, Germany. The VLBA is an instrument of the National Radio Astronomy Observatory, a facility of the National Science Foundation operated under cooperative agreement by Associated Universities, Inc. 
\end{acknowledgements}

\bibliographystyle{aa}
\bibliography{aa_abbrv,bibliography}

\end{document}